# THE ANGULAR CORRELATION FUNCTION OF GALAXIES IN CDM AND CHDM MODELS

G. Yepes

Departamento de Física Teórica C-XI, Universidad Autónoma de Madrid, Cantoblanco 28049, Madrid SP

A.A. Klypin[1]

Department of Astronomy, New Mexico State University, Las Cruces NM 88001

A. Campos and R. Fong

Department of Physics, University of Durham, South Road, Durham DH1 3LE UK

## ABSTRACT

We estimate the angular correlation functions, $w(\theta)$, for the standard CDM, tilted $n = 0.7$ CDM and hybrid (CHDM) models, and compare with observations. When compared with the APM observational results scaled to the Lick depth, there appears to be fair agreement with the estimate from the CHDM model. But a more detailed comparison using the unscaled APM data for the five magnitude slices with $\Delta b_j = 0.5$ shows that, in fact, none of the models can actually fit $w(\theta)$ for all the slices simultaneously. The CDM and tilted CDM models are then seen to fall below the APM data by a factor $\sim 2$ at $\theta = 1°$, where the observed $w(\theta)$ is still relatively large ($\sim 0.1$) and, thus, probably less affected by any possible residual systematic errors. As $n = 0.7$ tilted CDM and SCDM bracket all possible tilted models, we conclude that none of the tilted models is consistent with the APM results. With CHDM, a $w(\theta)$ amplitude that is 30%–40% too high on scales $\theta < 0°.5$ is predicted for the deepest slices ($b_j \approx 20$).

We find that no reasonable simple variation of the parameters for the luminosity function or for the evolution of the correlation function with redshift could change the situation. Thus, the discrepancy between the APM data and the CHDM model, though small, seems to be real.

*Subject headings:* cosmology: theory – dark-matter – large-scale structure of universe –galaxies: clustering

## 1. INTRODUCTION

The standard CDM model (SCDM) (Blumenthal *et al.* 1984; Davis *et al.* 1985) has proved to be a very attractive theory. But, despite its many successes, there seems now to be a growing body of observational evidence challenging the viability of the model. It would seem that it is the shape of the SCDM power spectrum that is basically wrong, regardless of its normalization. For review of CDM problems see Ostriker (1993) and Liddle & Lyth (1993). Several variations of the SCDM model have, thus, been proposed. All these modifications of the SCDM model have a free parameter that can be tuned to match simultaneously the observational constraints imposed by COBE (Smoot *et al.* 1992) and by the APM results for the large-scale clustering of galaxies (Maddox *et al.* 1990, MESL). Among the most frequently considered models

[1] Also Astro-Space Center, Lebedev Physical Institute, Profsojuznaja ul. 84/32, 117810 Moscow, Russia

are tilted CDM (e.g. Cen *et al.* 1992), CDM+Λ (i.e. a flat, low density CDM model with a cosmological constant: Efstathiou, Sutherland & Maddox 1990, Turner 1991, Gorski, Silk & Vittorio 1992, Cen, Gnedin & Ostriker 1992, Kofman, Gnedin & Bahcall 1993) and hybrid or CHDM models (CHDM have a combination of Cold and Hot dark matter components: Valdarnini & Bonometto 1985, van Dalen & Schaeffer 1992, Davis, Summers & Schlegel 1992, Klypin *et al.* 1993, Nolthenius, Klypin & Primack 1994, Cen & Ostriker 1994).

The galaxy angular correlation function, $w(\theta)$, has become one of the classical tests for cosmological models. As only photometry is involved, two-dimensional sky catalogues of galaxies can probe much deeper and more completely than three-dimensional ones. Although the clustering level is smaller due to projection, this is compensated by the very much larger number of objects in such sky catalogues. Furthermore, analysis of the projected distribution avoids the effects and uncertainties caused by the distortions in redshift space due to the peculiar velocities of galaxies. The best estimates to date of $w(\theta)$ to large angular scales come from the APM galaxy survey (MESL), containing $\sim 2$ million galaxies with $b_j \leq 20.5$.

Comparisons with APM has usually been done for the APM data scaled to the depth of the shallower Lick catalogue. For the tilted CDM model with slope $n = 0.7$, Cen *et al.* (1992) claim that $w(\theta)$ fits well the APM data scaled to the Lick depth, but no results were explicitly shown. A similar result was also claimed for the CDM+Λ model with $0.1 \leq \Omega h \leq 0.3$ (Efstathiou, Sutherland & Maddox 1990; Kofman, Gnedin & Bahcall 1993). However, the CDM+Λ models have too steep a slope for the 3D correlation function, $\xi(r)$, with $\gamma > 2$ at small $r$ (Davis *et al.* 1985, Gramann 1988), leading to a gross overestimate of $w(\theta)$ at small angles. For example, it is a factor of two larger than the APM $w(\theta)$ at $\theta = 0°\!.1$ ( Efstathiou, Sutherland & Maddox 1990). This effect was not taken into account by Kofman, Gnedin & Bahcall (1993), who assumed the same shape for $\xi$ as that of SCDM. For the CDM+Λ model to be compatible with observations, galaxies in the model would need to be very antibiased with respect to dark matter on scales $\lesssim 1$ Mpc. But, nobody has yet been able to demonstrate this. The angular correlation function for the CHDM model has been estimated by Jing *et al.* (1993) only for the depth of the Lick catalog. Their results indicate that a fit to the APM $w(\theta)$ should not be a problem for the CHDM model.

Information that has not been much exploited is that of the APM estimates of $w(\theta)$ for five magnitude slices with $\Delta b_j = 0.5$. Compared to just comparing results for $w(\theta)$ at the Lick depth, we can, in principle, explore two important factors for theoretical estimates for the slices: possible evolution of the correlation function, $\xi(r,z)$, and the evolution of the luminosity function. However, for the magnitude range under consideration, the results appear to be not very sensitive to any reasonable evolution in $\xi$. As for the second factor, it affects the amplitude of $w(\theta)$, but does not change its shape. And it is the shape of the APM $w(\theta)$ which cannot be explained by the SCDM model for the formation of structure in the universe.

In this letter we present then a detailed comparison of the clustering predicted by the SCDM, tilted CDM and hybrid models with the most recent observational estimates of $w(\theta)$ obtained using the APM catalog, for which the residual errors of Maddox, Efstathiou & Sutherland (1990) have been taken into account.

## 2. MODELS

To find the 'galaxy' two-point correlation function, $\xi$, for the SCDM and CHDM models, we use the results of N-body simulations to estimate it at short scales and the predictions of linear theory for large scales. The CHDM model is for $\Omega_{\rm cold} = 0.6$, $\Omega_{\rm hot} = 0.3$ and $\Omega_{\rm baryons} = 0.1$, and has the initial fluctuation spectrum given by Klypin *et al.* (1993), which is normalised to the COBE quadrupole of $17\mu K$ (Smoot *et* al. 1992), corresponding to a linear bias of $b = 1.5$. The same biasing parameter is also used for normalising the SCDM power spectrum. For the $n = 0.7$ tilted CDM model, we have simply adopted the results of Cen et al. (1992), who give the biasing parameter as $b = 2$.

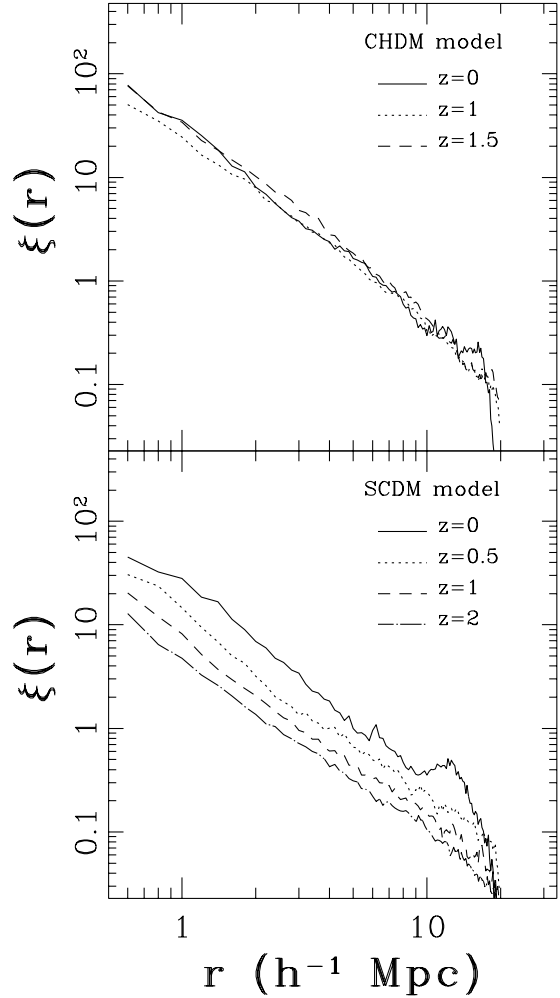

Fig. 1.— Two-point spatial correlation function for the dark halos in the N-body simulations for the Hybrid model (CHDM) and for the Standard CDM model with $b = 1.5$, as a function of the comoving distance and for different redshifts. $\xi(r)$ has been computed weighting each dark halo by its mass to avoid the overmerging problem in the simulations.

The simulations were carried out using a particle-mesh code on a $512^3$ mesh in a 100 Mpc box ($H = 100h$ km s$^{-1}$ Mpc$^{-1}$, $h = 0.5$) with $256^3$ cold particles and, additionally for the CHDM simulation, $2 \times 256^3$ hot particles. Dark halos ('galaxies') are identified as the local maxima of the total density defined on the $512^3$ mesh. The mass assigned to a 'galaxy' is the mass inside the cell where the maximum is found. Only dark halos corresponding to $\Delta\rho/\rho \geq 50$ are used to estimate $\xi(r,z)$. The number of dark halos in the simulations is 14635 and 27210 in the CHDM and SCDM models respectively. To estimate the correlation function for the *luminous* matter in the simulations we have weighted each dark halo by its mass. This is done to compensate for the artificial 'overmerging' of dark halos due to the lack of resolution. A more detailed description of the simulations and results can be found elsewhere (Nolthenius, Klypin & Primack 1994, Klypin, Nolthenius & Primack 1994). However, the initial power spectra of the models analysed here had extra power for the longest waves in the simulation box, a statistical fluke due to the small number of harmonics at these wavelengths. Because the waves are still in the linear regime even at $z = 0$, we can, for the CHDM simulation, simply subtract the contribution of the extra power from the correlation function using linear theory. A comparison with a new CHDM simulation, which did not have the fluke, shows that the extra power slightly affected small scales. To compensate for this effect, we reduced the correlation length, $r_0$, by 8% and reduced $\xi$ on larger scales by 15%. The extra power had little impact on the SCDM simulation and, so, no corrections were needed in this case.

Figure 1 shows the correlation functions of dark halos in comoving coordinates for the SCDM and the CHDM models at different redshifts. In the SCDM model, the $z-$dependence of the correlation function can be fitted as $\xi(x,z) \sim \xi(x, z=0)/(1+z)^{3/2}$, where $x$ is the comoving distance. In proper coordinates $\xi(r)$ grows as $\propto (1+z)^{-3.2}$ for $z \lesssim 2$. This is consistent with the classical solution $\xi(r) \propto (1+z)^{-3}$ for stable clustering (Phillipps et al. 1978, Peebles 1980). As is well known, in the SCDM model clustering proceeds hierarchically, with small galaxy clusters forming first and with these merging to form larger clusters. Thus, as the crossing time is much less than the Hubble time, these clusters are virialized or close to being so and we would expect them to be stable. However, we see here that in CHDM $\xi(r)$ is not stable in proper space, but is essentially constant in comoving space. It looks as if the clustering is 'frozen-in', i.e. is expanding with the universe. But halos are still forming as well as merging. Thus, it would seem that this model predicts the 'similarity' growth of structure.

Defining $\xi(r)$ as the correlation function at $z = 0$, we have, thus, assumed the simple form of $\xi(r,z) = \xi(r) \cdot (1+z)^{\epsilon}$, with $\epsilon = -3.2$ for the SCDM model and $\epsilon = -1.7$ for CHDM. For the tilted CDM model, we have also assumed the same dependence on $z$ as that for the SCDM model. Because of the finite box size, $\xi(r)$ becomes negative at $20h^{-1}$Mpc, which, of course, is our reason for using linear theory to estimate $\xi$ on large scales. Although a power-law slope of $\gamma \approx 1.8$ fits best the numerical results, this would not provide a good fit to the APM data, which requires a value of 1.7. As this is not excluded by our results, when account is taken of the possible uncertainties in our estimates, a power-law slope of $\gamma = 1.7$ has, thus, been assumed for all the models. That is, at radii smaller than some limit $x_{\mathrm{lin}}$, $\xi(r) = (r/r_0)^{-\gamma}$, where $r_0$ is the correlation length, and for $r > x_{\mathrm{lin}}/(1+z)$ we use the results of the linear theory scaled up by $b^2$, where $b$ is the biasing parameter. $x_{\mathrm{lin}}$ is, thus, the position at which the power-law crosses over the linear theory result. Note that $\xi(r)$ becomes negative at some point, $r_c$, which is $31h^{-1}$Mpc for SCDM and $50h^{-1}$Mpc for CHDM. Although the amplitude of the negative part of $\xi(r)$ is small, $\sim -2.5 \times 10^{-3}$, we shall see in § 3 that it contributes significantly to the estimate of $w(\theta)$. For very large scales, we use the asymptotic form $\xi(r) \sim -r^{-3}$ for all models. The correlation lengths (at $z = 0$) from the simulations are $r_0 = 5.5h^{-1}$Mpc and $r_0 = 6.2h^{-1}$Mpc for SCDM and CHDM, respectively, and that for the tilted CDM model is $5.0h^{-1}$Mpc (Cen *et al.* 1992). The transition between the power-law behavior and the linear approximation are at $x_{\mathrm{lin}} = $ 24, 13.5, and 8 $h^{-1}$ Mpc for the CHDM, tilted CDM and SCDM models, respectively.

## 3. RESULTS FOR $w(\theta)$ AND DISCUSSION

We calculate the angular correlation functions for the models using the full relativistic Limber's formula (Phillipps et al. 1978, Eq. 13), with the selection function determined using the evolving galaxy luminosity function of MESL: $\phi(x) \propto x^{\alpha} \exp(-x)$, $x \equiv 10^{0.4(M^* - M)}$, $M^* = M_0^* + M_1^* z$, $\alpha = \alpha_0 + \alpha_1 z$, $M_0^* = -19.8$, $M_1^* = 1$, $\alpha_0 = -1$, $\alpha_1 = -2$.

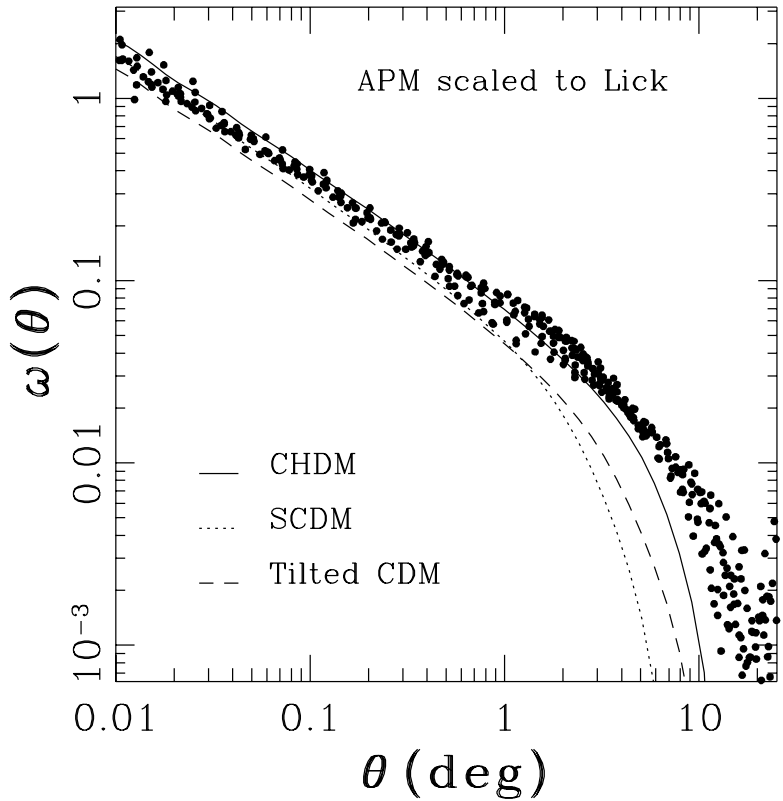

Fig. 2.— Angular correlation function for the APM data scaled to the depth of the Lick catalog (dots). The curves represent the predictions for $w(\theta)$ of the three models considered: SCDM, tilted CDM ($n = 0.7$) and CHDM. The same luminosity function evolution with redshift has been used to rescale the APM data and the models.

Figure 2 presents the APM $w(\theta)$ data scaled to the Lick depth and the predictions for the three models. As expected, SCDM fails to match the observed $w(\theta)$ for $\theta \gtrsim 1°$. With the tilted CDM model, the theoretical curve is systematically lower than the observational one. It could be argued that the amplitude of $\xi(r)$ has been underestimated. However, a shift of the curve to match the data at short scales would clearly not be enough to reproduce the observed large scale power. As $n = 0.7$ tilted CDM and SCDM bracket all possible tilted models (Cen et al. 1992), we conclude that none of the tilted models is consistent with the APM results. As for the CHDM result, it is similar to the result of Jing et al. (1993) for one of their hybrid models and provides the best of the model fits. However, there is still a clear underprediction at $\theta > 3°$. This may seem surprising, as the power law at small scales extends to $24h^{-1}$Mpc, whereas the APM 'break' is reported as being at $20h^{-1}$Mpc. However, this latter scale is based on a phenomenological two-power-law fit to the data, whereas here the negative part of the model $\xi(r)$ is clearly having a very significant effect.

It is even more interesting to actually compute model $w(\theta)$ for the six narrow APM magnitude slices of $\Delta b_j = 0.5$ between $b_j = 17.34$ and 20.52. Figure 3 presents the APM data for 3 of these slices as well as the predictions of the models; the results for the other slices are in between these. As before, both SCDM and tilted CDM models are unable to fit the APM data. But, for the CHDM model there is a clear qualitative difference from the comparison with the APM data scaled to the Lick depth as in Figure 2. The CHDM model is in good agreement with the observational data at large angular scales for the deep slices, but underestimates $w(\theta)$ in the shallow ones. Furthermore, at short scales the model now fits well the data of the shallow slices, but overestimates them in the deep ones. This was lost in the previous comparison, because, when scaled to the Lick depth, the data for the different slices cross such that at short scales the upper data points (i.e. those closer to the model predictions) correspond to the shallow slices, whereas at large scales the lower points (now the closest to the model) are those corresponding to the deep slices. Thus, fitting to the slices themselves shows that the CHDM model here seems also not to provide a satisfactory fit to the APM $w(\theta)$.

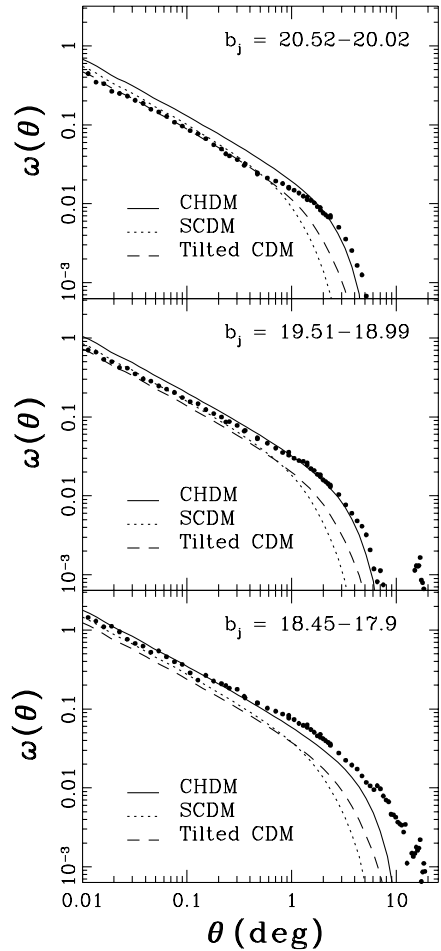

Fig. 3.— Angular correlation function for 3 magnitude slices of the APM catalog and the corresponding predictions of the theoretical models: CHDM (solid line), tilted CDM (dashed line) and SCDM (dotted line). These data correspond to the new and yet unpublished version of the APM survey. The data have been corrected for residual errors and are slightly different as compared to the original version (MESL).

It could be argued that this discrepancy has arisen because we have overestimated $r_0$ for the model. However, adjusting $r_0$ to make the CHDM model fit $w(\theta)$ at small scales for the deep slices would also make it fail to fit the data at large angles. A similar effect would happen if we were to change $\gamma$. A slope of $\gamma = 1.6$ gives a very good fit to the APM data for the deep slices, at both small and large angular scales, but then makes the fit to the shallow slices even worse.

Table 1: Dependence of $\omega(\theta)$ on parameters (see §3)

| Model, Params. | Angle | | | | | | | |
|---|---|---|---|---|---|---|---|---|
| | $0°.1$ | | $1°.0$ | | $3°.0$ | | $5°.0$ | |
| | slice-1[†] | slice-2[†] | slice-1 | slice-2 | slice-1 | slice-2 | slice-1 | slice-2 |
| APM | 0.3180 | 0.0890 | 0.0750 | 0.0154 | 0.0250 | 0.0045 | 0.0120 | 0.0006 |
| CDM MESL | -0.25 | -0.14 | -0.56 | -0.54 | -0.78 | -0.95 | -0.91 | -1.29 |
| CHDM MESL | +0.09 | +0.45 | -0.22 | +0.25 | -0.20 | -0.13 | -0.27 | +0.01 |
| CHDM $\epsilon = -3$ | -0.03 | +0.16 | -0.30 | +0.05 | -0.25 | -0.07 | -0.26 | +0.73 |
| CHDM No Evol | +0.22 | +0.67 | -0.13 | +0.45 | -0.09 | +0.08 | -0.14 | +0.43 |
| CHDM $\alpha_1 = 0$ | -0.05 | +0.15 | -0.33 | -0.05 | -0.34 | -0.46 | -0.43 | -0.70 |
| CHDM $M_1 = -1$ | -0.19 | -0.14 | -0.44 | -0.32 | -0.46 | -0.69 | -0.56 | -1.01 |

[†] Slice-1: $b_j = 18.45 - 17.90$; Slice-2: $b_j = 20.52 - 20.02$

To test the effects of possible uncertainties in the luminosity function and in how $\xi(r)$ may evolve, we varied some of these parameters. Table 1 summarizes our results. We present the 'errors' of $w(\theta)$ in the form: Error = $(w_{\text{model}} - w_{\text{APM}})/w_{\text{APM}}$, at four representative angles, $0°.1, 1°, 3°$ and $5°$, with the APM $w(\theta)$ in the first row of the Table for reference. We changed only one parameter at a time, keeping the rest the same, to probe the effect of just that parameter on $w(\theta)$. A non-evolving Schechter luminosity function, with a $k-$correction of $3z$ has also been used. It can be seen from Table 1 that this gives similar results as for the MESL luminosity function, but this is not surprising as MESL's aim was to obtain a number-redshift, $n(z)$, distribution of galaxies that had a similar form to that of a no-evolution model. In detail, the $w(\theta)$ estimate is higher than the observations when the no evolution model is used. It is about 15% higher for all the slices at short scales and is always within a factor of 2 for $\theta < 10°$. The differences are larger for the deeper slices. By changing the rate of evolution to $\xi \propto (1+z)^{-3}$, we can improve the CHDM results for the deep slices at short scales, but then the problem at large angles ($\theta > 3°$) increases and gets significant.

Thus, if there are no systematic errors of any significance remaining in the APM $w(\theta)$ estimates, more theoretical work is needed still to find even a viable model for how large-scale structure may have formed in the Universe. Nevertheless, it is not easy to find a reasonable candidate among existing cosmological models. For example, the primeval baryon isocurvature (PBI) model (Peebles 1987, Cen, Ostriker & Peebles 1993) can provide enough power on large scales to match $w(\theta)$ on large angles. But the slow rate of growth of fluctuations in an open Universe at low redshifts will probably result in an even larger (as compared to CHDM) $w(\theta)$ for deep slices, making the fit even worse.

Unfortunately, there is still the problem of possible residual errors in the zero-points for the APM survey (Metcalfe, Fong & Shanks 1994). In effect, this could mean that the form of $n(z)$ is quite uncertain for the APM data. It would seem then that a more robust way of comparing the model predictions with the data is simply to choose a reasonable selection function so as to fit $w(\theta)$ at short scales, where also the signal is higher. A better fit at short scales is achieved when the luminosity function evolution becomes flatter (Table 1). However, not surprisingly, we have now exacerbated the fit at large angular scales. But, it is still possible that the APM $w(\theta)$ at these scales may also be significantly affected by residual zero-point errors (Fong, Hale-Sutton & Shanks 1992, Fong, Metcalfe & Shanks 1994). The main conclusion that we can then draw from our analysis is that none of the current cosmological models provide a good fit to the observed angular distribution of galaxies as given by the APM data, although claims have been made that the CHDM model could provide a good fit.

We would like to thank Steve Maddox for kindly providing us with the new version of the APM data on $w(\theta)$ prior to publication. GY wishes to thank DGICyT (Spain) for financial support under project number PB90-0182.